\documentclass[11pt,conference,compsoc,a4paper,onecolumn,twoside]{IEEEtran}

\usepackage{amsmath}
\usepackage{graphicx}
\usepackage{geometry}
\usepackage{vmargin}
\usepackage[T1]{fontenc}
\usepackage[latin1]{inputenc}
\usepackage{afterpage}
\usepackage{listings}
\usepackage{subfigure}
\usepackage{url}
\usepackage{picins}

\makeatletter
\let\@ORGmakecaption\@makecaption
\long\def\@makecaption#1#2{\@ORGmakecaption{#1}{#2}\vskip\abovecaptionskip\relax}
\makeatother

\setmargins{3.0cm}{3.0cm}       
           {15.0cm}{23.7cm}     
           {0pt}{0pt}         
           {0pt}{30pt}          

\begin{document}
\bstctlcite{myctlfullname}

%
 \renewcommand{\thesection}{\arabic{section}}
 \renewcommand{\thesectiondis}{\thesection.}
 \renewcommand{\thesubsectiondis}{\thesectiondis\arabic{subsection}.}
 \renewcommand{\thesubsection}{\thesection.\arabic{subsection}}
 \renewcommand{\thesubsectiondis}{\thesubsection.}
 \renewcommand{\thesubsubsectiondis}{\thesubsectiondis\arabic{subsubsection}.}
 \renewcommand{\thesubsubsection}{\thesection.\thesubsection.\arabic{subsubsection}}

\title{Integrated monitoring of multi-domain backbone connections\\
\Large{Operational experience in the LHC optical private network}}

\author{
\IEEEauthorblockN{
Patricia Marcu, David Schmitz, Wolfgang Fritz, \\Mark Yampolskiy, Wolfgang Hommel}
\IEEEauthorblockA{
Leibniz Supercomputing Centre\\
Boltzmannstr. 1, 85748 Garching, Germany\\
\{marcu,schmitz,fritz,yampolskiy,hommel\}@lrz.de}
%
%
%
%
}

\maketitle

\begin{abstract}
\textit{Novel large scale research projects often require cooperation between various different project partners that are spread among the entire world. They do not only need huge computing resources, but also a reliable network to operate on. The Large Hadron Collider (LHC) at CERN is a representative example for such a project. Its experiments result in a vast amount of data, which is interesting for researchers around the world. For transporting the data from CERN to 11 data processing and storage sites, an optical private network (OPN) has been constructed. As the experiment data is highly valuable, LHC defines very high requirements to the underlying network infrastructure. In order to fulfil those requirements, the connections have to be managed and monitored permanently. In this paper, we present the integrated monitoring solution developed for the LHCOPN. We first outline the requirements and show how they are met on the single network layers. After that, we describe, how those single measurements can be combined into an integrated view. We cover design concepts as well as tool implementation highlights.}


\end{abstract}


\section{Introduction}\label{sec:1intro}

%
%
%
%

The significant increase in the availability of high-speed research networks has led to the deployment of large-scale distributed computing environments that are able to serve a great number of geographically separated users by exchanging huge amounts of data. Direct communication between sites and computing facilities is now necessary in many working environments. This results in greatly expanded requirements for high-speed, dedicated networks that cross multiple domains. The European Research Network G\'EANT \cite{Gea10} and National Research and Educational Networks (NRENs) in Europe are high-capacity networks, which are based on optical technologies and components that provide wavelength-based services to the research community. 

A representative project is the provisioning of the networking infrastructure for the Large Hadron Collider (LHC) at CERN in Switzerland. Its research experiments produce about 15 petabytes of data per year. Therefore, a multi-domain LHC Optical Private Network (LHCOPN) was established \cite{BMM05}, dedicated to support data exchange. The LHCOPN consists of Tier-0 and Tier-1 centres connected by End-to-End (E2E) links. These E2E links connect organisations (Tier-1 centres) that are located in different countries and cross the shared network infrastructure of different providers towards the Tier-0 centre at CERN. 

One of the most important and difficult issues related to this dedicated network is network management. The monitoring and troubleshooting of optical networks and individual E2E links is challenging. Researchers all over the world are increasingly using dedicated optical paths to create high-speed network connections, and different groups of users may want to use monitoring applications for specific research purposes. They need access to network measurement data from multiple involved network domains, visualise network characteristics and troubleshoot related issues \cite{hbbd05a}. 

A quick overview and visualisation of the network status is necessary to establish demarcation points that help distinguish network issues within LHCOPN from those in the sites. The deployment of monitoring tools and the use of common services should provide a unified network information view across all domains \cite{haya07}. Typical off-the-shelf solutions do not provide such functionality. 

This article is structured as follows: In Section \ref{sec:2req}, the requirements in the context of the LHCOPN are described. After that, some important existing approaches in the area of inter-organizational monitoring are outlined in Section \ref{sec:3sota}. Section \ref{sec:4monLayer12} explains, how monitoring is done at layer 1 and 2, before monitoring at layer 3 and above is added in Section \ref{sec:5monLayer3}. A motivation of integrating them and a description of this integration is outlined in Section \ref{sec:6integrView}. Section \ref{sec:7operAndImpl} then gives an overview of the implementation and the operational experiences gained within the LHCOPN. The article is concluded in Section \ref{sec:9concl}. 


\section{Requirements}\label{sec:2req}

The monitoring of the LHCOPN, i.e.\ Tier-0 and Tier-1 centres, and the links between them poses several new challenges: 

\paragraph{Multi-domain monitoring}

The LHCOPN itself is based on resources that are supplied by several academic networks such as G\'EANT, European NRENs, Internet2, ESnet and Canarie. Therefore, a solution has to be found to collect monitoring data from all these self-administered networks to form a joint view of the resulting network.

\paragraph{Monitoring of different layers}

While academic networks have been used to monitor the network layer, the LHCOPN requires E2E links on the data link layer to be monitored. These are based on heterogeneous technologies, as the different participating networks use different technologies. E2E links are formed by combining technologies such as SDH/SONET, native Ethernet or Ethernet over MPLS, where each domain is dependent on the data that it can retrieve from the network management system of its vendor.

\paragraph{Joint view of all metrics}
In the visualisation a view has to be formed by combining E2E link and IP-related monitoring data and by linking these data in a suitable manner. In doing so, it must be considered that there are also several data sources on the IP level, in particular the retrieval of SNMP data from routers and the results of active measurements.

\section{Existing approaches}\label{sec:3sota}

The issue of multi-domain monitoring is not only a challenge in the context of the LHCOPN, but also in the general operation of networks. In 2004 a collaboration of the GN2/GN3 project with Internet2, ESnet, RNP and others was started to jointly develop a communication protocol and tool set under the name perfSONAR  \cite{hbbd05, perfSONAR}. This development has become necessary by the limitations of existing tool sets which were tied to single domain monitoring and limited to subsets of metrics that can be monitored. Such limitations apply e.g.\ to the MonALISA  \cite{MonALISA} tool set. Apart from being used in the LHCOPN, the perfSONAR tools are used within the networks that participate in the collaboration. As perfSONAR is open source, it is used by other projects like Distributed European Infrastructure for Supercomputing Applications (DEISA) \cite{deisa} too.

There are already several tools which can visualise perfSONAR measurement data \cite{hjkm06}. One of them is perfsonarUI which can be used for troubleshooting by allowing a direct interaction with perfSONAR measurements.  

All principles of the perfSONAR protocol will be taken into account in the customisation for the LHCOPN, especially its \textit{multi-domain-monitoring} feature. Furthermore, this will be done with respect to the \textit{different layers monitoring} also developed within the GN2/GN3 projects. The missing requirement for this customisation is the \textit{joint view on all metrics}. Also, a global overview of the LHCOPN was needed, in which different layer views coexist. This customisation was achieved by a dedicated version of the Customer Network Management (CNM) tool  \cite{CNM1} (in a browser-based version).

\section{Monitoring at layer 1 and 2: E2EMon}\label{sec:4monLayer12}

%
%

The introduction of hybrid networks and the possibility to deliver E2E links that involve multiple domains has led to the need to monitor these links. E2E Links are defined as dedicated optical point-to-point network connections realized at ISO/OSI layers 1 and 2. For the purpose of fault detection and localization of such connections, a dedicated tool called E2EMon (E2E Monitoring Tool) \cite{haya08a} has been developed over the recent years. It is used by the E2E Coordination Unit (E2ECU) in G\'EANT -- an organizational unit established for inter-domain coordination of operational procedures.


E2EMon basically consists of two important components---the Monitoring Point (MP) and the central component. Every domain which provides a segment of such an E2E link needs to have an E2EMon Measurement Point (MP) in place which retrieves data from the local network management system to provide status information for the link segment. In the following, we will first outline, how the Measurement Points works, continuing with a detailed description of the central component.

\subsection{E2EMon measurement point (MP)}

\begin{figure}[t]
\centering
\includegraphics[width=0.6\textwidth]{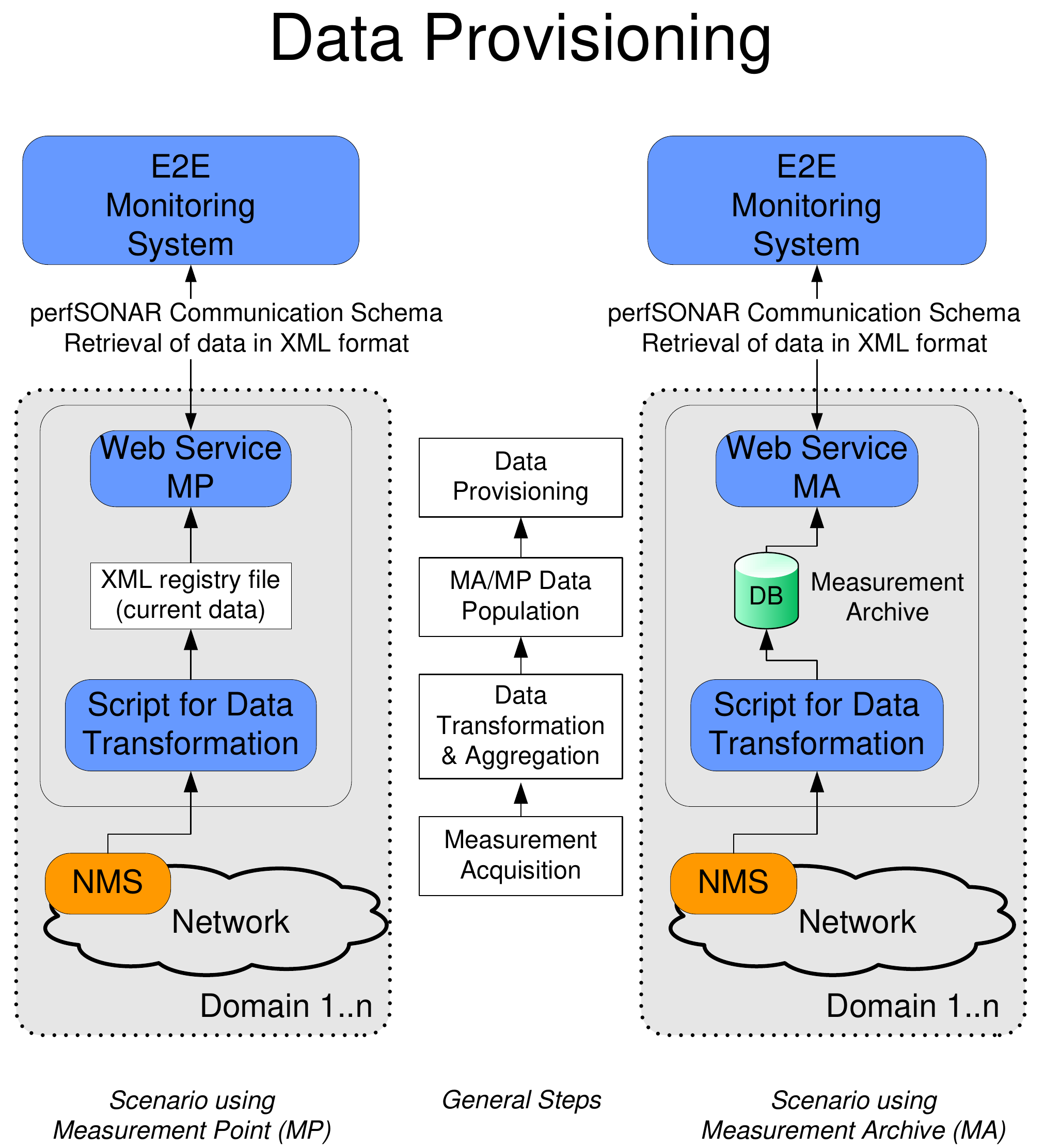}
\caption{Functionality of E2EMon MP \cite{haya08a}}
\label{fig:e2emonmp_provisioning}
\end{figure}

The Monitoring of circuits crossing different organizational domains is often a big challenge. It is not quite easy to get all the needed information from all involved router interfaces and so on. Furthermore, many network operations centres (NOCs) already have their own local monitoring solution, which they want to keep. Therefore, an infrastructure and technology independent collection of this local monitoring data is needed. The E2EMon MP provides an interface between those provider specific and a provider independent representation of monitoring information. 

The E2EMon MP is a set of perl scripts. Once installed and deployed in every involved domain, it is basically a standalone SOAP server instance. Therefore, it does not rely on webservers like tomcat, apache, etc. It does not install itself as a daemon service, but of course can be run as such if it is configured in the operating system. All necessary and required modules are installed together with the service.

After that, E2EMon MP is waiting for an XML file to send to the central component. This XML file has to be generated by the particular network monitoring system (NMS) in use, for example, Nagios, Cacti, or others (see Figure \ref{fig:e2emonmp_provisioning}).

Obviously, this XML file has to be updated regularly---otherwise it will not reflect the actual situation. The XML file is based on the schema introduced by the OGF Network Measurements Working Group (NMWG) \cite{NMWG, SZG04}. A typical snippet of such an XML file is shown below in Listing 1.

\small
\begin{center}
\begin{minipage}[b]{120mm}
\lstset{language = XML}
\begin{lstlisting}[caption=Sample E2EMon MP XML file,frame=tlrb]{}
<nmwg:message type="SetupDataRequest"
      xmlns:nmwg="http://ggf.org/ns/nmwg/base/2.0/">
   <nmwg:metadata id="meta1">
     <nmwg:eventType>Path.Status</nmwg:eventType>
   </nmwg:metadata>
   <nmwg:data id="data1" metadataIdRef="meta1"/>  
</nmwg:message> 
\end{lstlisting}
\label{lst:sample_xml_mp}
\end{minipage}
\end{center}
\normalsize

The XML file itself does not contain specific detailed measurement data, but rather an abstracted view of it. The interfaces and links are categorized as UP, DOWN, DEGRADED, or UNKNOWN, depending on their actual operational state. This mapping has to be provided by the domain's monitoring system, respectively by the tool or script that converts the local monitoring system's data to XML. Last but not least, the XML file consists of different sections, each providing information about either a so called \emph{monitored link} or so called \emph{demarcation points} needed by the central component of the E2E Monitoring System . We will explain these two terms in detail in the next section. 

\begin{figure}[t]
	\centering
		\includegraphics[width=0.75\textwidth]{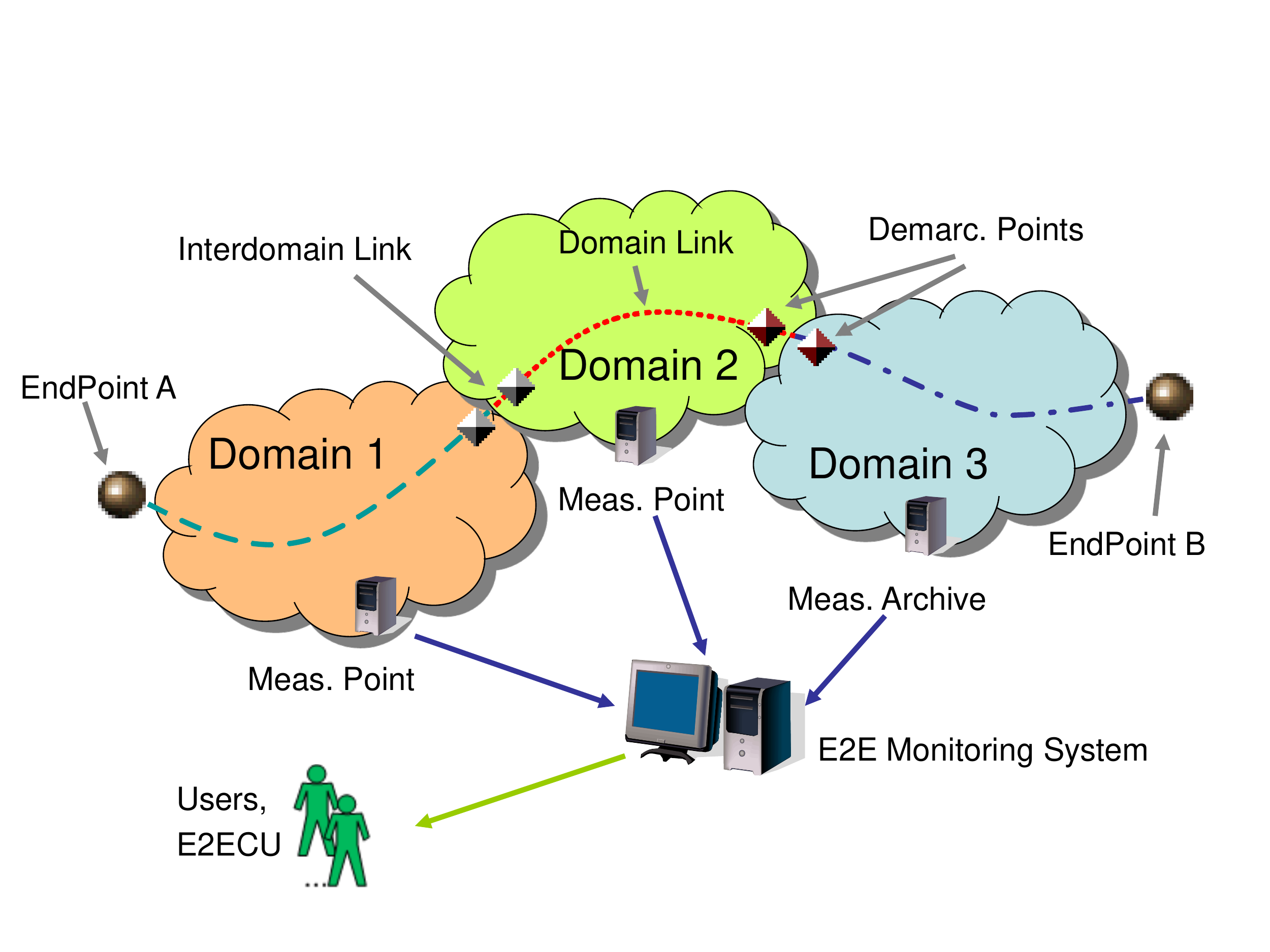}
	\caption{Information flow: from domains to E2EMon and E2ECU \cite{haya08a}}
	\label{fig:E2Emon_fig1_E2ELink_Structure}
\end{figure}

\subsection{E2E link Monitoring System (E2EMon)}

E2EMon relies on the ,,health'' information provided by different NRENs. A term ,,Monitored Link'' is used by E2EMon developers and users in order to refer to E2E Link parts, for which monitoring information is provided by NRENs. All monitored links are edge-to-edge connections from a border of one NREN to the border of the same or neighbor NREN. 
Ends of Monitored Links are referred to as Demarcation Points (DPs). 
The Following three types of Monitored Links are supported by E2EMon:

\begin{itemize}
\item \textbf{Domain Link} is an connection completely realized within a single NREN
\item \textbf{InterDomain Link} is a whole connection, interconnecting two neighboring NRENs
\item \textbf{InterDomain Link Part} is the type of connection representing a part of Inter-Domain  connection from the perspective of a single NREN
\end{itemize}

The distinguishing between InterDomain Link and InterDomain Link Part is necessary because of typically very restrictive information and management policies of the involved NRENs. In most cases these policies prevent the measurement of inter-domain connection state. Therefore, ,,InterDomain Link Part'' is used to provide state information from a single NREN's perspective. States of two parts can be aggregated to the state of the whole inter-domain connection. Also, the restrictive information policies cause high abstraction levels of Monitored Links.

One of the challenges, which have to be overcome in a multi-domain environment is the synchronization of monitoring data. E2EMon does not require clock synchronization in multiple NRENs. Instead, E2EMon polls MPs/MAs of different NRENs periodically (see Figure \ref{fig:E2Emon_fig1_E2ELink_Structure}). E2EMon assumes then that data retrieved from multiple NRENs in the same polling cycle is up to date and synchronized. Currently, a 5 minute polling interval is used. This time interval is treated by E2ECU as fine grained enough. The time needed to collect and to process data from about 30 NRENs is about 1 minute. 

\begin{figure}[t]
	\centering
		\includegraphics[width=0.75\textwidth]{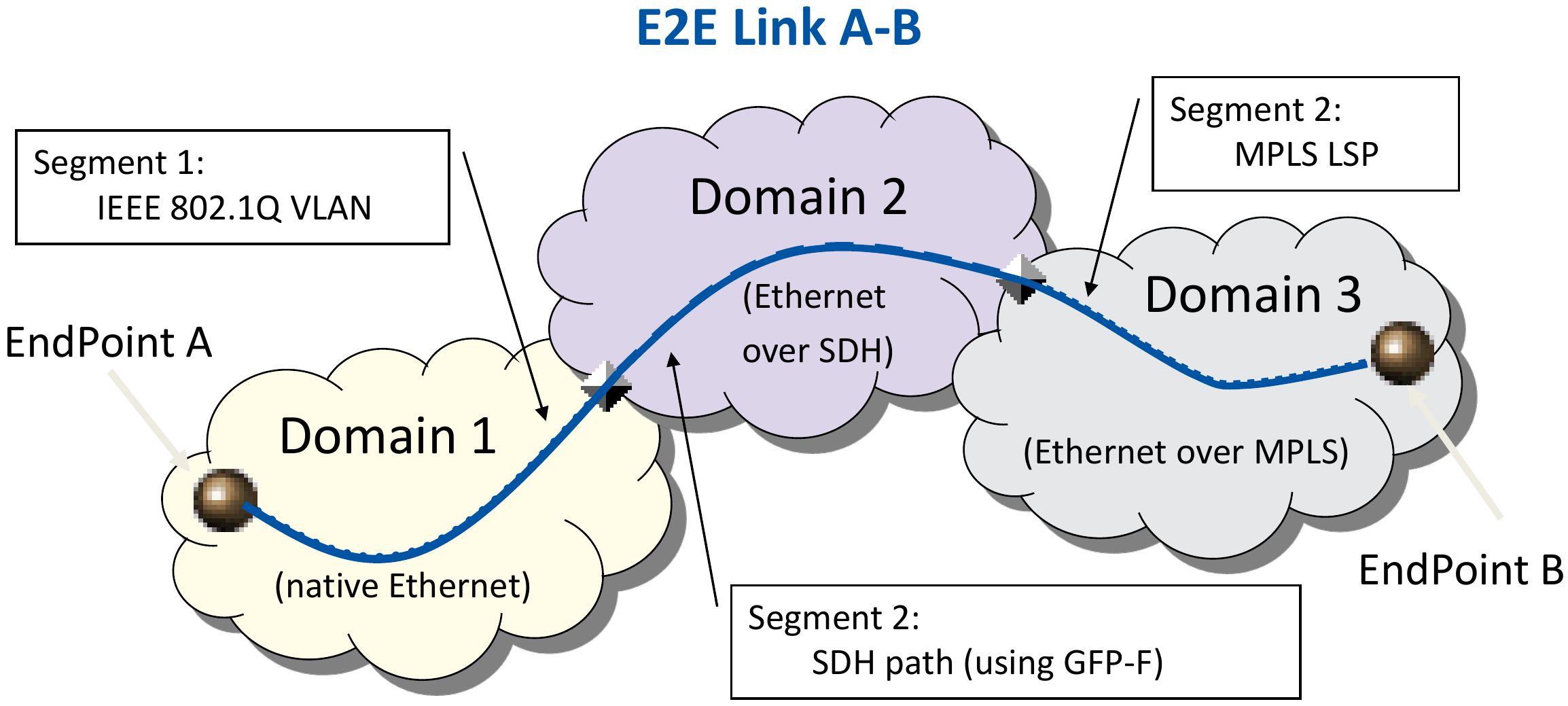}
	\caption{E2E Link, typical built up \cite{yamp09}}
	\label{fig:E2Emon_fig2_E2ELink_Realisation}
\end{figure}

\begin{figure}[h]
\centering
\subfigure[Operation States]
 {\includegraphics[width=0.315\textwidth]{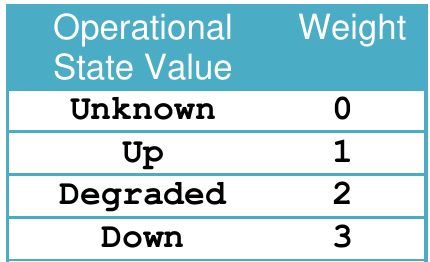}\label{tab:E2Emon_tab1_OperationalStates}}
\subfigure[Administrative States]
 {\includegraphics[width=0.67\textwidth]{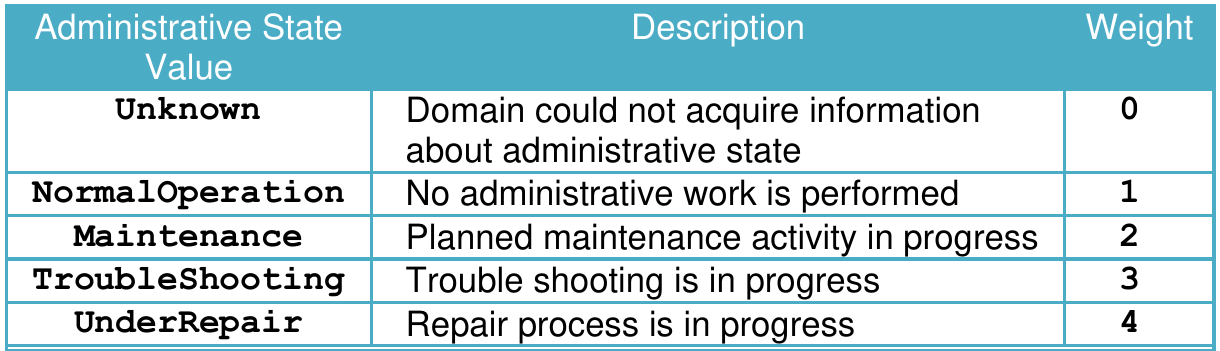}\label{tab:E2Emon_tab2_AdministrativeStates}}
\caption{E2EMon States \cite{haya08a}}
\end{figure}

Because of the independence of NRENs and non-coordinated procurement policies, heterogeneous hardware and network technologies are used in general to realize E2E Links parts. Typical are SDH, Ethernet, Ethernet over SDH, and MPLS (see Figure \ref{fig:E2Emon_fig2_E2ELink_Realisation}). Due to high heterogeneity and as an outcome of this a high difference of state information, a true technology specific monitoring data cannot be used for multi-domain monitoring. Instead, the following abstracted operational states are used:
\begin{itemize}
\item \textbf{UP} -- the connection is up and running  
\item \textbf{DEGRADED} -- the connection is up, but with degraded performance
\item \textbf{DOWN} -- the connection is down and cannot be used at all
\item \textbf{UNKNOWN} -- the state of the connection is unknown 
\end{itemize}
Identical aggregation rules have been defined for (a) two InterDomain Link Part states to the state of a whole InterDomain Link and (b) all Monitored Links of a particular E2E Link to the whole E2E Link state. The aggregation rules are defined in the way, that the worst state dominates. By implementating such a strategy, values have been assigned for all supported states (see Figure \ref{tab:E2Emon_tab1_OperationalStates}). In the aggregation rule, the biggest state-weight of involved Monitored Link is used as a weight of the whole E2E Link. If one or more Monitored Link states are UNKNOWN, the GUI shows a warning independent of the aggregated state.

In addition to the operational state, E2EMon supports an administrative state of monitored and E2E links. The supported states, their weight and short description are listed in Figure \ref{tab:E2Emon_tab2_AdministrativeStates}. The rule to aggregate the administrative state is identical to the one used for the operational state.  The administrative state reflects the management processes performed by the domains. This state is used by E2EMon in order to, e.g., prevent a new alarm from being raised if the Monitored Link goes DOWN during a planned maintenance. 


In practice, some of the described features of E2EMon are not used. As there is no strict definition of semantics for DEGRADED state, NRENs generally omit to report it even if the degraded performance is recognized. Another unused feature is the possibility that an NREN reports the whole state of an Inter-Domain connection. In order to do this, an access to the infrastructure of a neighbor NREN is needed. But as this collides with the restrictive security policies of some NRENs, only reports of Domain Link Parts are used currently. Finally, the administrative state is used by few NRENs only. The coupling of NREN-specific management tools and the export of the data for E2EMon is difficult. 
The state of E2E Links is presented in the E2EMon GUI. All E2E Links are shown in a semi-graphical view, which provides information about the Monitored Links and their order in the link. E2EMon does not require information about order and arrangement of Monitored Links in advance. Instead, E2EMon computes this information implicitly as follows:

\begin{itemize}
\item For every Monitored Link the globally unique E2E Link ID is provided, which allows to group information provided by different NRENs. 
\item Every Monitored Link is further specified by globally unique IDs of its Demarcation Points. 
\end{itemize}

\begin{figure}[htbp]
	\centering
		\includegraphics[width=0.65\textwidth]{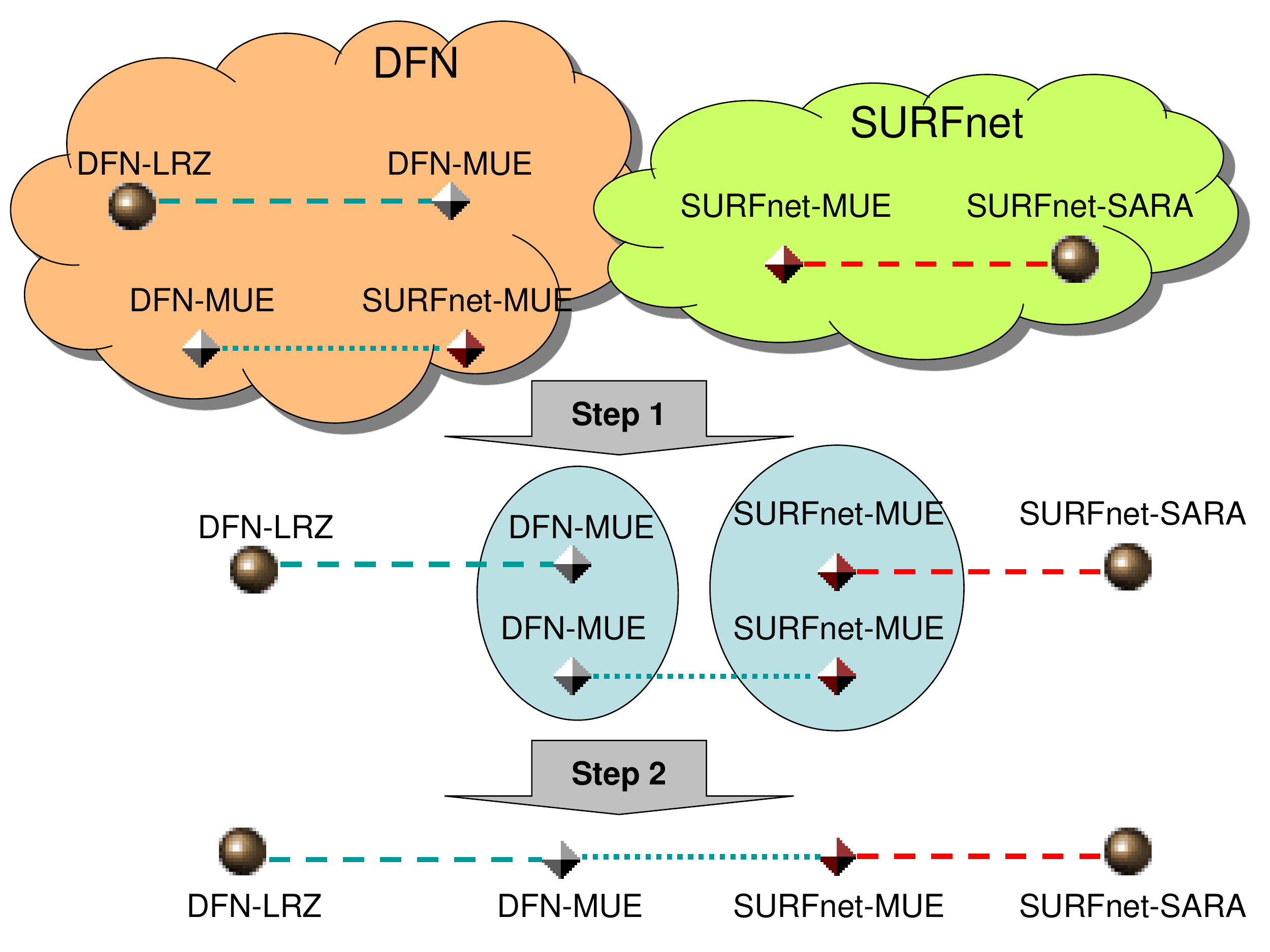}
	\caption{Reconstruction of an E2E Link \cite{haya08a}}
	\label{fig:E2Emon_fig3_E2ELink_Reconstruction}
\end{figure}

Based on this information, different Monitored Links are ,,stitched together'' at the DPs with the same IDs (see Figure \ref{fig:E2Emon_fig3_E2ELink_Reconstruction}).

\begin{figure}[htbp]
	\centering
		\includegraphics[width=1.00\textwidth]{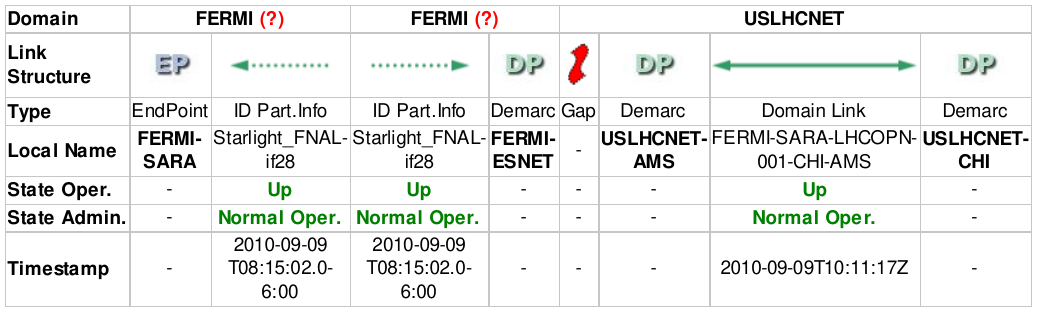}
	\caption{GUI representation of an E2E Link \cite{e2emon}}
	\label{fig:E2Emon_fig4_Screen_E2ELink_Structure}
\end{figure}

In some cases, the reconstruction of the whole E2E Link structure might not be possible. This can happen, e.g., if some MPs/MAs can not be reached due to firewall restrictions. It is also possible that information about monitored links is provided using wrong IDs for DPs. Such situations sometimes occur if new E2E Links are setup. In those cases, E2EMon assembles as many pieces of an E2E Link as possible and displays them with the icons for gaps between contiguous sections (see Figure \ref{fig:E2Emon_fig4_Screen_E2ELink_Structure}). 

\begin{figure}[htbp]
	\centering
		\includegraphics[width=1.00\textwidth]{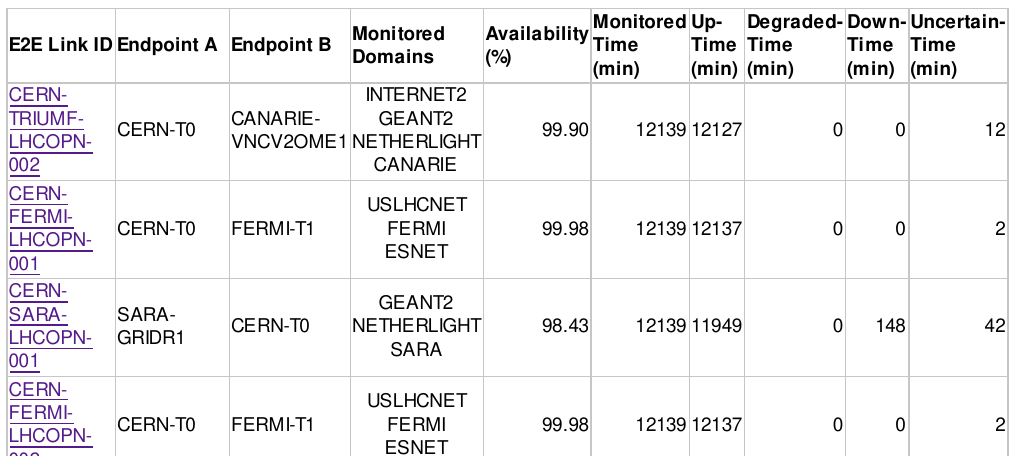}
	\caption{Monthly statistics of productive E2E Links \cite{e2emon}}
	\label{fig:E2Emon_fig5_Screen_Statistics}
\end{figure}

E2EMon distinguishes between productive and not yet productive E2E Links. For productive E2E Links, monthly and weekly availability statistics are collected. These statistics contain not only UP and DOWN time of a particular E2E Link, but also the so called ,,Uncertain-Time''. If the E2E Link structure could not be fully reconstructed, the state of the link is calculated based on the information of known monitoring links. At the same time, this state is considered as ,,uncertain'' as it can be influenced by states of unknown monitored links. If an E2E Link is in uncertain state, the polling period is added to the link's uncertain time counter. The availability of an E2E Link is computed as certain Up-Time divided by the overall monitoring time of the link. Statistical information can be accessed via the GUI and also saved as a CSV (comma separated value) export file (see Figure \ref{fig:E2Emon_fig5_Screen_Statistics}).

E2EMon is a tool integrated in manual operational processes. Alongside with the GUI, several interaction interfaces have been implemented. In case an E2E Link state changes to DEGRADED or DOWN, E2EMon automatically sends email notifications to E2ECU and involved NRENs. In order to simplify the integration of E2EMon with other management tools like Nagios, SNMP traps are generated every time an E2E Link state changes. The state of all productive E2E Links is exported using an XML file. This XML export is updated every polling interval and used by LHCOPN Weathermap for integrating layer 1 and 2 states with other measurements.

\section{Monitoring at layer 3 and above: HADES and BWCTL}\label{sec:5monLayer3}

For IP-level monitoring, the HADES and BWCTL tools are of interest, as they provide relevant metrics and are integrated into the perfSONAR framework. Therefore, they were easily integrated as a part of the LHCOPN overall management solution. 

\subsection{HADES}\label{subsec:hades}

HADES (Hades Active Delay Evaluation System) \cite{hades} uses dedicated hardware boxes to perform active tests in the network to measure delay, jitter, packet loss and traceroute (with respect to IPPM recommendations \cite{rfc2330}). For precise timing, GPS antennas are installed in addition to the hardware boxes. Networking Time Protocol (NTP) can also be used but with less precision.

The HADES \cite{hades} boxes have been deployed at the Tier-0 and Tier-1 LHCOPN locations to provide QoS measurements.
The HADES topology is made up of the abstract nodes, so it can easily link them to E2E measurements and directed (uni-directional) \textit{HADES links} between them. A HADES link is determined by its source and targeted abstract nodes. As HADES links correspond to pairs of abstract nodes, they are identified by ordered pairs of abstract locations. 

HADES measurements are run as a full mesh between all nodes. To prevent users from being overloaded with too much data, the visualisation in the integrated view (the so-called \emph{Weathermap}, see Section \ref{sec:6integrView}) is limited to measurements that relate to paths where E2E links exist. 

The metrics on the HADES layer are IP performance metrics (IPPM) computed for each HADES link (one way delay \cite{RFC2679}, IP delay variation (jitter) \cite{RFC3393} and packet loss \cite{RFC2680}) as well as the hop list/count metric. As the links are directed between two different HADES end points A and B, the metrics exists for $A\rightarrow B$ and $B\rightarrow A$. All these metrics have a time resolution of 5 minutes. The HADES metrics are stored in a HADES Measurement Archive (based on the SQL MA), which is used to store and publish historical monitoring data produced by the HADES Measurement Points.

\subsection{BWCTL}\label{subsec:bwctl}

Similar to HADES, BWCTL (Bandwidth Test Controller) \cite{bwctl} verifies available bandwidth from each endpoint to other points to identify throughput problems. In the LHCOPN, BWCTL nodes are included within the HADES boxes by using a second interface card.
 
Each BWCTL end point address is associated with an abstract node. This is a 1:1 mapping but the IDs of the BWCTL end point and of the abstract node are not the same (BWCTL IP addresses vs. location names).

On this layer, the needed metrics are minimum, medium and maximum BWCTL throughput (stored in the SQL MAs). As the BWCTL links are directed between two different BWCTL end points A and B, these BWCTL metrics exist for both directions $A\rightarrow B$ and $B\rightarrow A$. 

\section{Integrated View: LHCOPN Weathermap}\label{sec:6integrView}

Network operators rely on monitoring information. Different layers provide different information of the monitored links. In order to foster the operational procedures, an integrated view of the monitoring information is needed. To provide such an integrated view, all the measurements important to the LHCOPN have to be identified and a mechanism has to be defined how they can be combined.

\subsection{Measurements in LHCOPN}\label{subsec:measure}

\begin{figure}[ht]
\centerline{\includegraphics[width=0.75\columnwidth]{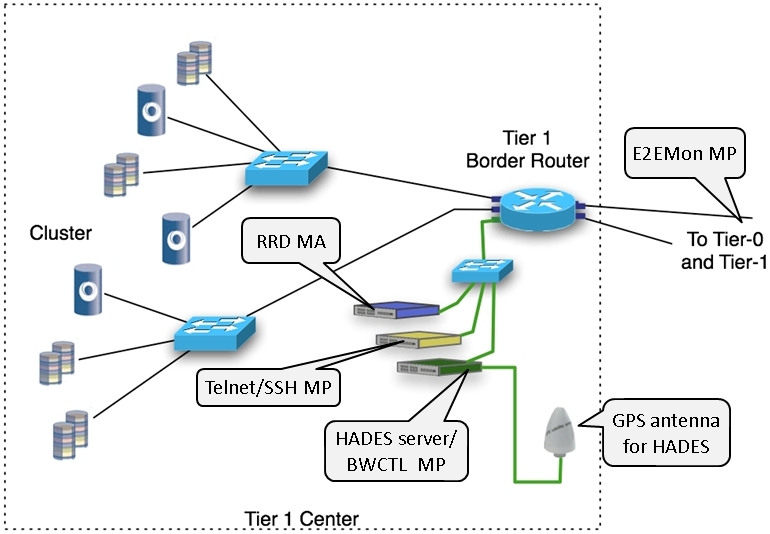}}
\caption{Tier-1 site configuration}
\label{fig:measurLHCOPN}
\end{figure}

To understand the metrics displayed in the LHCOPN Weathermap, it is necessary to know how the measurements are carried out. The deployment of the perfSONAR measurement tools at each Tier-1-centre is therefore shown in Figure \ref{fig:measurLHCOPN}.

First of all, data is collected for the E2E links (see Section \ref{sec:4monLayer12}). The links start at the Tier-0 centre or at one of the Tier-1 centres (backup links), end at one of the Tier-1 centres and typically cross several administrative domains  (e.g.\ G\'EANT, European NRENS, Internet2 or ESnet). The status of each link is then calculated based on the NMS data from each domain involved. \\
In addition to this E2Emon MP, three measurement servers are located at each centre: 
\begin{itemize}
	\item \textbf{HADES/BWCTL MP:} The first server is the HADES box, which conducts one-way delay \cite{RFC2679}, IP delay variation (jitter) \cite{RFC3393}, packet loss \cite{RFC2680}, and traceroute tests with any other HADES box in the LHCOPN every minute. It is connected to a GPS antenna for precise timing. In addition, the HADES box hosts a BWCTL MP which is used for throughputs with the Tier-0 centre every 8 hours. The BWCTL MP is hosted on a different interface to avoid interference with HADES measurements.  
	\item \textbf{Telnet/SSH MP:} The second server is used for the Telnet/SSH MP, a tool that allows configuration data to be retrieved from the routers. It is only mentioned here for completeness, but does not carry out any regular measurements. 
	\item The last server is used to host an \textbf{RRD MA} to collect utilisation, interface errors and output drop data related to the router located at the Tier-1 centre. 

\end{itemize}
%
%
%

\subsection{Integrating the monitoring data}
\label{subsec:integr}

The measurements that are carried out have to be displayed in a suitable manner, which means in this case that a trade-off between correct display and usability has to be made. For example, HADES measurements are not directly located on the routers, so that delay data is not exactly measured at the location of utilisation measurements. Events on the short link between router and HADES box can lead to wrong interpretations. 

Even more difficult considerations have to be made for E2E link status data and its relation to IP metrics. By default the IP data in the network uses the direct way via an E2E link. However, if the E2E link fails (including the optical protection), then the IP protection performs a rerouting. 
Although IP packets are still transferred, they take another physical route. Therefore, it is necessary to clearly distinguish between optical and IP level.  

For this reason, a data model is introduced in the following that covers all topology information per network layer and all necessary topology mapping information for the LHCOPN Weathermap.

With respect to the requirements stated in Section \ref{sec:2req}, the operators of the LHCOPN need a global view on their network. Other essential aspects are layer-related, location-related and metric-related views on the LHCOPN. An E2E view is needed to check the availability of the E2E links involved. Therefore, as described in the former sections different layers (topologies) have been defined: E2E link, HADES and BWCTL. 



Information about the IP links between two different IP interfaces is needed to determine the status of the links between two IP interfaces within the LHCOPN (these are VLANS in their terminology). 

The router topology consists of the abstract nodes, which here relate to IP interfaces and \textit{IP links} (pairs of IP interfaces). One IP link corresponds to a VLAN in the LHCOPN terminology. The current assumption is that one abstract link is associated with one IP interface pair only. This means that, if one abstract link has two or more E2E links, they both contribute (in an aggregated manner) to the same IP link (back-up link or bundle of links). Also, one E2E link can contribute to a single VLAN only. 

The metrics used on the router topology are utilisation, input errors and output drops for each end point of an IP link. These metrics have a time resolution of 5 minutes. Typically, they are retrieved via SNMP from routers and stored in so called RRD MAs (Round Robin Database Measurement Archive) or SQL MAs (Structured Query Language Measurement Archive). These are tools that provide archived measurement data.

\section{Operational experience and implementation highlights}\label{sec:7operAndImpl}

\begin{figure*}[t]
\centerline{\includegraphics[width=0.80\textwidth]{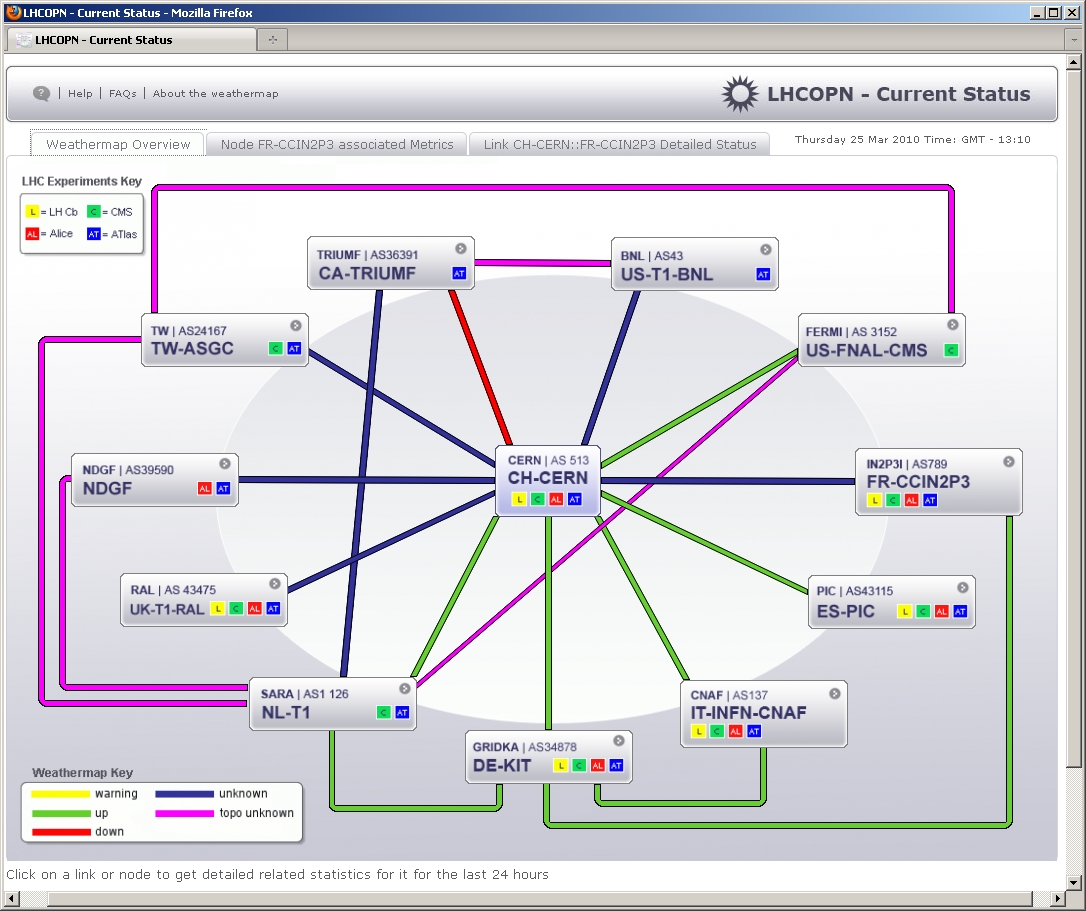}}
\caption{A view on the E2E Link Topology Tab in the LHCOPN Weathermap tool}
\label{fig:overview}
\end{figure*}

To satisfy the LHCOPN requirement for multi-domain monitoring accessed through a global view, a layer based on the E2E link has been specified to 
form the main view. This layer gives an overview of the whole LHCOPN, on the dedicated E2E links involved in the LHCOPN. For each
link that is displayed in the topology two kinds of abstractions can be involved. A link represented here can be an E2E Link or it can be
an E2E link plus another E2E link which serves as optical (1+1) protection. The other kind of abstraction that is involved, is that for each E2E link
the status is derived from data retrieved from multiple NMS. In the following a detailed description of the E2E Link topology, whose representation in the LHCOPN Weathermap is shown in Figure \ref{fig:overview}, is given.

The topology consists of \textit{abstract nodes} and \textit{abstract links}. Abstract nodes represent the Tier-0 and Tier-1 LHCOPN locations and are named accordingly. They abstract the exact location where measurements are conducted in order to allow easy linking of this topology
to the other topologies. The abstract links are non-directed (i.e. bidirectional) links between the abstract nodes.  

The metric used for this abstract layer is the \textit{aggregated status} for each abstract link. It is computed every 5 minutes from the E2EMon status of all associated E2E links. For a single E2E link the status is retrieved in the E2EMon system by polling all E2EMon MPs every 5 minutes.

\subsection{Data retrieval, filtering and integration}\label{subsec:data_retriev}
The data retrieval is concerned with the fetching and updating of topology information, topology mapping information, metric mapping information (see Sections \ref{sec:4monLayer12} and \ref{sec:5monLayer3}), as well as the actual metric data fetching. The measurements in LHCOPN as well as metric data fetching were outlined in Section \ref{subsec:measure}.


The \textit{abstract topology} and its associated E2E links have to be imported from a structured, static configuration file provided by LHCOPN users or, in the future, from an online configuration file. 


The \textit{E2E Link topology} is fetched from the E2EMon export interface together with their individual E2E link states and have to match the ones (associated with abstract links) specified above. 

The \textit{HADES topology} is imported from metadata of the LHCOPN HADES MA. The topology mapping (abstract node 1:1 HADES node) is trivial, and has to be altered to show links that are interesting to the Weathermap (links that correspond to an abstract link). 

The \textit{BWCTL topology} is imported from metadata in the LHCOPN BWCTL SQL MA. The mapping between BWCTL nodes (BWCTL IP addresses) and abstract nodes is statically configured.

Potential LHCOPN IP interface address pairs are imported from the metadata of various LHCOPN RRD MAs and then need to be altered according to the abstract link to IP interface address pair mappings specified above. 

%



\begin{figure}[h]
\centering
\subfigure[The link status tab]
 {\includegraphics[width=0.49\textwidth]{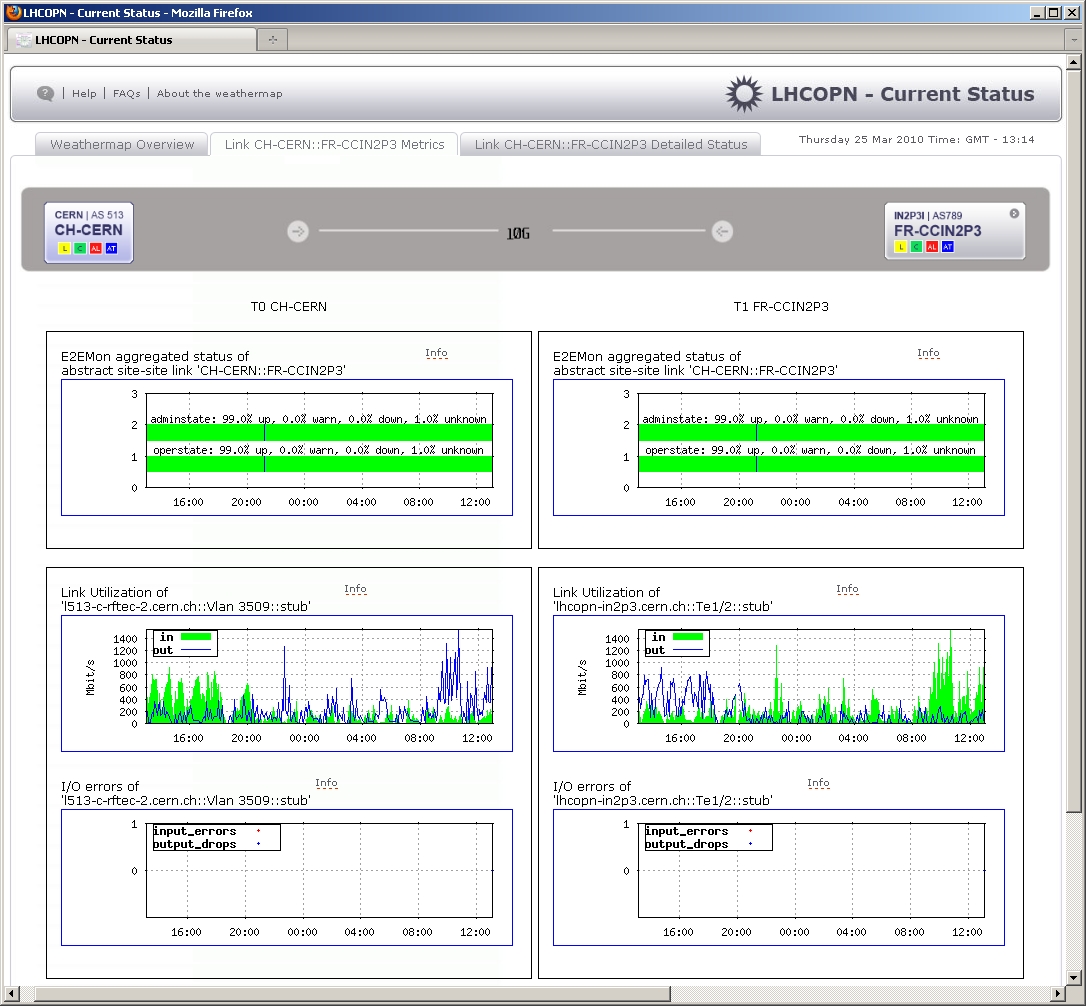}\label{fig:linkstatus}}
\subfigure[The node status tab]
 {\includegraphics[width=0.49\textwidth]{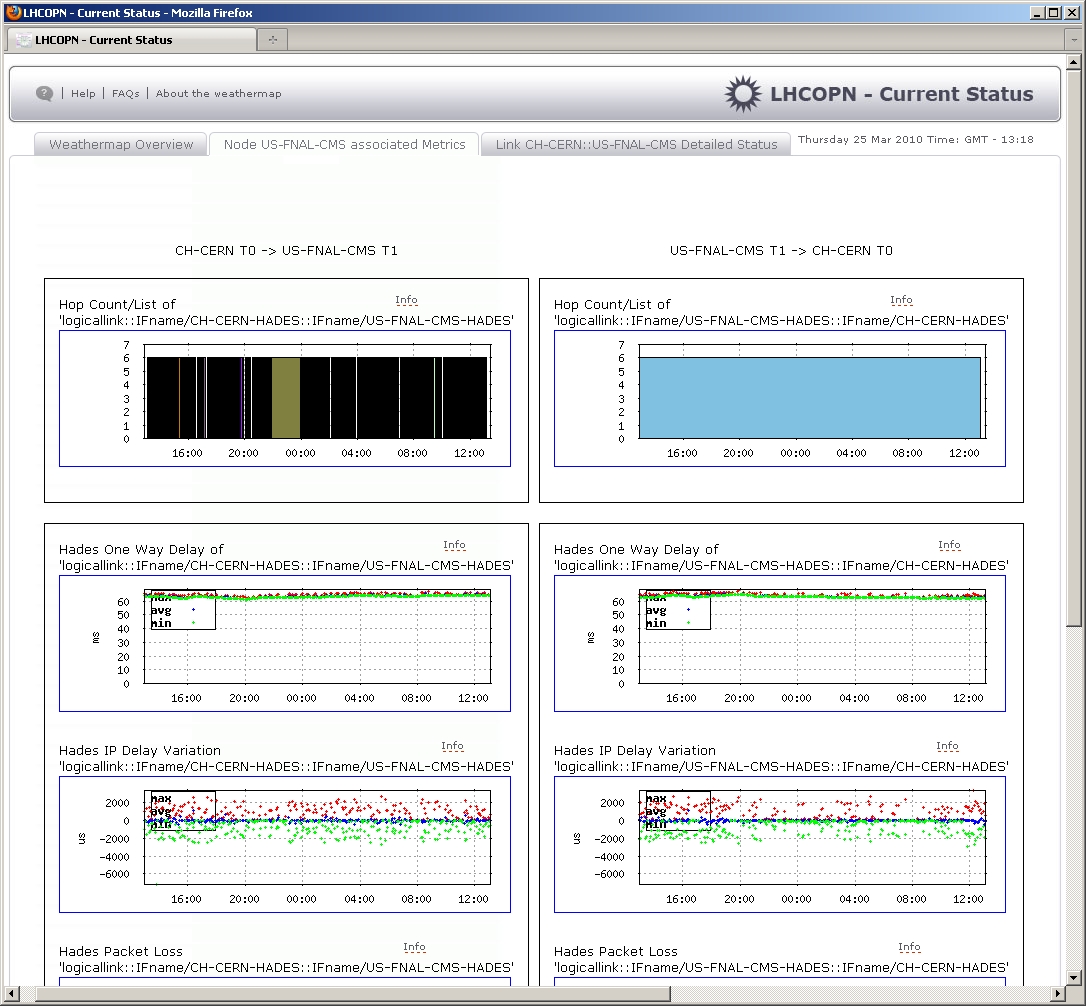}\label{fig:nodestatus}}
\caption{The link and status tabs}
\end{figure}

\subsection{Visualisation}\label{subsec:visual}

To meet the requirements of LHCOPN users, a visualisation consisting of three tabs has been designed: \textit{Overview Map Tab}, \textit{Metric Tab} and \textit{E2E Link Tab}.

\subsubsection{The Overview Map Tab}\label{ovmap}

In this tab (see Figure \ref{fig:overview}) a map consisting of abstract nodes and abstract links (described in Section \ref{sec:5monLayer3}) is shown. This overview map indicates the current status using four colours: RED, YELLOW, GREEN and BLUE for the current abstract link status DOWN, WARNING, UP and UNKNOWN  (defined in Section \ref{sec:5monLayer3}). 




In addition to that, the fifth status (MAGENTA) does not represent a value of the metric's aggregated status, but instead indicates that there is a serious mismatch in the topology mapping concerning the abstract link; namely that all associated E2E links (from the point of view of the abstract topology) are unknown to the E2E topology (from the point of view of the E2E topology). This status is called topology unknown and indicates that no aggregated status for the abstract link could be computed. 


The content of the other two tabs (Metric tab and E2E Link tab) is shown when clicking either on an abstract link or an abstract node in the map. So by clicking and choosing the selected abstract element, data corresponding to this abstract link or node is loaded in the metric tab and E2E link tab. 

\subsubsection{The Metric Tab}

The metric tab shows statistical graphs of metrics associated to particular abstract links.


If an abstract link in the overview map is selected, data for this specific link is shown. If a Tier-1 abstract node is selected, the abstract link from the Tier-0 abstract node (CERN) to this selected Tier-1 abstract node location is selected. If the Tier-0 abstract node (CERN) is selected, data for all abstract links from CERN to any Tier-1 is displayed.

In the metric tab, 24-hour metric graphs of various metrics of the network layers (see Sections \ref{sec:4monLayer12} and \ref{sec:5monLayer3}) are presented for the chosen abstract link. The list of visualised metrics is different depending whether the selected abstract element is a node or a link. 




\textbf{\textit{Metrics for an abstract link:}}
Selecting an \textit{abstract link} in the overview map displays the following metrics for this link (see Figure \ref{fig:linkstatus}) in the Metric tab:

\begin{itemize}
 \item The graph of the E2E aggregated status associated with the abstract link itself. This is based on the data model in Section \ref{subsec:data_retriev} and is visualised in the previously mentioned status colours.
 \item The RRD MA metrics graphs (see Section \ref{subsec:integr} for the single IP link associated with the abstract link. These are visualised for both IP interfaces at both end points of the IP link.
\end{itemize}

All the metrics are measured and updated every 5 minutes.

\textbf{\textit{Metrics for an abstract node:}}
Selecting an \textit{abstract node} in the overview map (all Tier-0 to Tier-1 abstract links related to the selected abstract node) displays statistic graphs of the following metrics for the chosen abstract links (see Figure \ref{fig:nodestatus}):


\begin{itemize}


\item The \textit{Hop count metric graph} is divided into differently coloured areas, indicating different routes. 
\item The \textit{HADES metrics} are visualised as scatter plot graphs (values are dots), each with a 5 minute time resolution. For one way delay and jitter the minimum, medium and maximum is needed.
\item \textit{BWCTL metric graphs} are visualising the minimum, medium, and maximum BWCTL throughput.

%

\end{itemize}

\begin{figure}[t]
\centerline{\includegraphics[width=0.55\columnwidth]{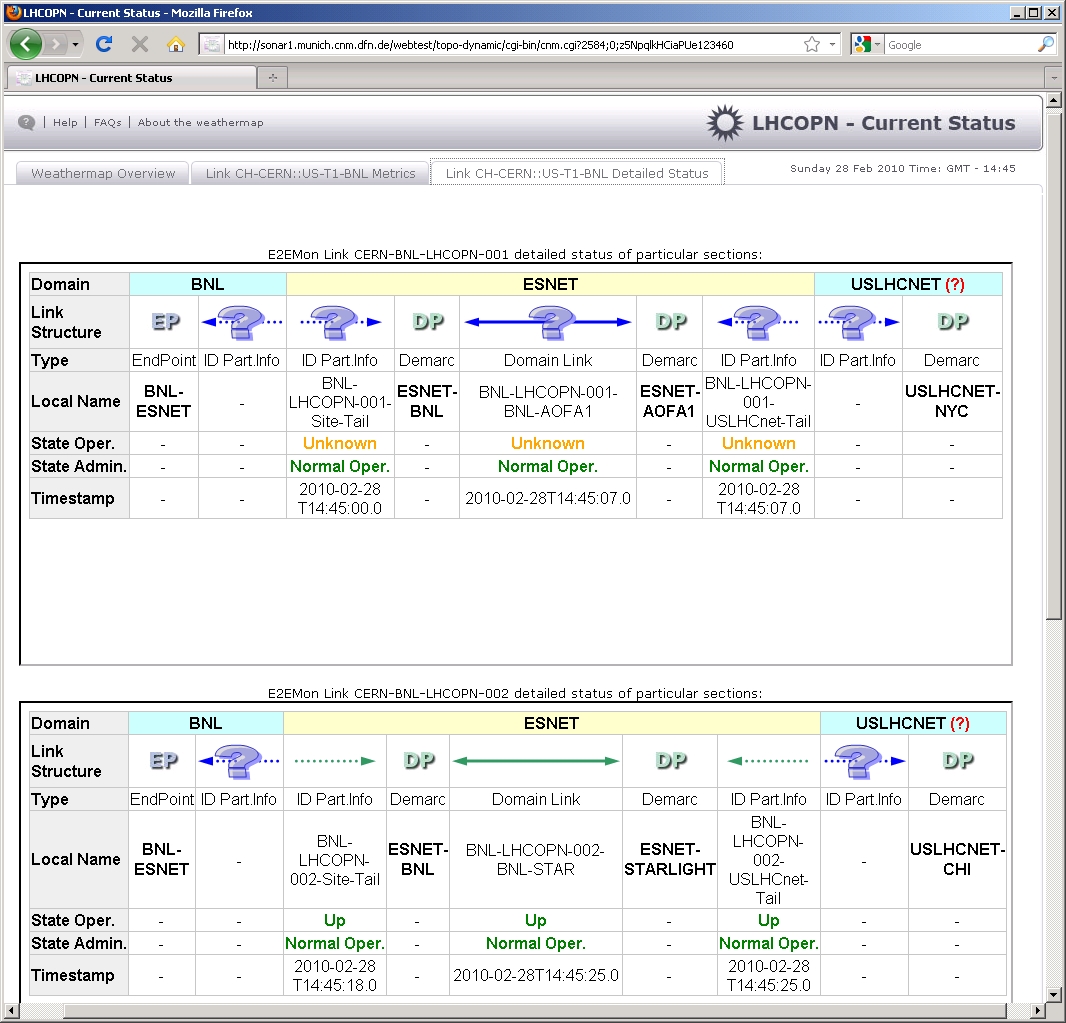}}
\caption{The E2EMon status tab}
\label{fig:e2emonstatus}
\end{figure}

\subsubsection{The E2E Link Tab}

The metric tab specified in the previous Section shows metrics on different (network) layers in a more end-to-end like fashion between the Tier-0/Tier-1 locations. In addition to this, the E2E link tab (see Figure \ref{fig:e2emonstatus}) presents a section status view for the focused abstract link.
This is done by wrapping the HTML page for the E2EMon section status for each E2E link associated to the focused abstract link in the map overview tab.


If the selected element in the map overview tab is an abstract link, the E2EMon segment status is shown for all E2E links associated to this. 

If the selected element in the map overview tab is an abstract Tier-1 node, the E2EMon segment status is shown for all E2E links associated with this focused abstract link.  


\subsection{Client and Access Point}\label{subsec:client}


The client is implemented as a dynamic HTML page with some java script code used for the tabbing interface.

To speed up the access, some graphical parts of the content are created and cached in advance:
\begin{itemize}
\item The current 24-hour statistic plot of any network element necessary as specified in Sections \ref{sec:4monLayer12} and \ref{sec:5monLayer3} are usually updated on a 5 minutes basis.
\item The overview map which includes the link status colour is updated every 5 minutes.
\item Internal to the HTML dynamic generation scripts, additional data base content caching is performed to speed up access further. 
\end{itemize}

\section{Conclusion and future work}\label{sec:9concl}

In this article, the monitoring of the LHCOPN has been explained with a focus on the LHCOPN Weathermap. The support structure is ready to fulfil its needs and has proven its usefulness in day-to-day operations since the LHC experiments are running, and large amounts of data are actually transferred via the network. 

Besides continuous improvements to the already existing tools, an alarm tool is currently under development. It is designed to be quite flexible in terms of alarm generation, to be suitable for different user needs. 

The perfSONAR services used for the LHCOPN and the Weathermap provide a good basis for the future large scale projects in Europe. A collection of such projects can be found in the roadmap of the European Strategy Forum on Research Infrastructures (ESFRI) [17]. For the Weathermap, this means that different ways of customisation to meet the needs of other projects are going to be investigated. For projects that want to use dynamic circuits, the perfSONAR group is already investigating suitable monitoring methods.

\small

\section*{Acknowledgements}
The authors would like to thank their colleagues at the Leibniz Supercomputing Centre of the Bavarian Academy of Sciences and Humanities (see \url{http://www.lrz.de/}) for helpful discussions and valuable comments about this paper.

The authors wish to thank the members of the Munich Network Management Team (MNM-Team) for helpful discussions and
valuable comments on previous versions of this paper. The MNM Team directed by Prof. Dr. Dieter Kranzlm\"uller and Prof.
Dr. Heinz-Gerd Hegering is a group of researchers at Ludwig-Maximilians-Universit\"at M\"unchen, Technische Universit\"at
M\"unchen, the University of the Federal Armed Forces and the Leibniz Supercomputing Centre of the Bavarian Academy of
Sciences and Humanities. See http://www.mnm-team.org/.



\normalsize
\bibliographystyle{IEEEtran}
\bibliography{promobib}

\subsection*{Autors}
\label{sec:shortBio}
\normalsize
\label{sec:PatriciaMarcu}
\parpic[r]{\includegraphics[width=0.2\textwidth]{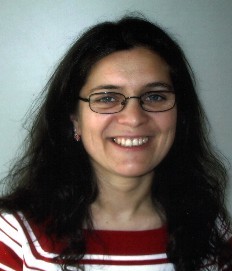}}
\textbf{Patricia Marcu} received diploma in Computer Science in 2006 at the Ludwig Maximilians Universität (LMU) München. In 2007 she joined the MNM Team at Leibniz Supercomputing Centre in Garching/Munich as a research assistant and pursues her Ph.D. degree in Computer Science. She is curently working on the further development Customer Network Managemnt (CNM) tool and on the visualisation of the LHCOPN within the european Géant projekt. Her research interest focuses on inter-oganisational fault management and IT Service Management.
\\

\label{sec:DavidSchmitz}
\parpic[r]{\includegraphics[width=0.2\textwidth]{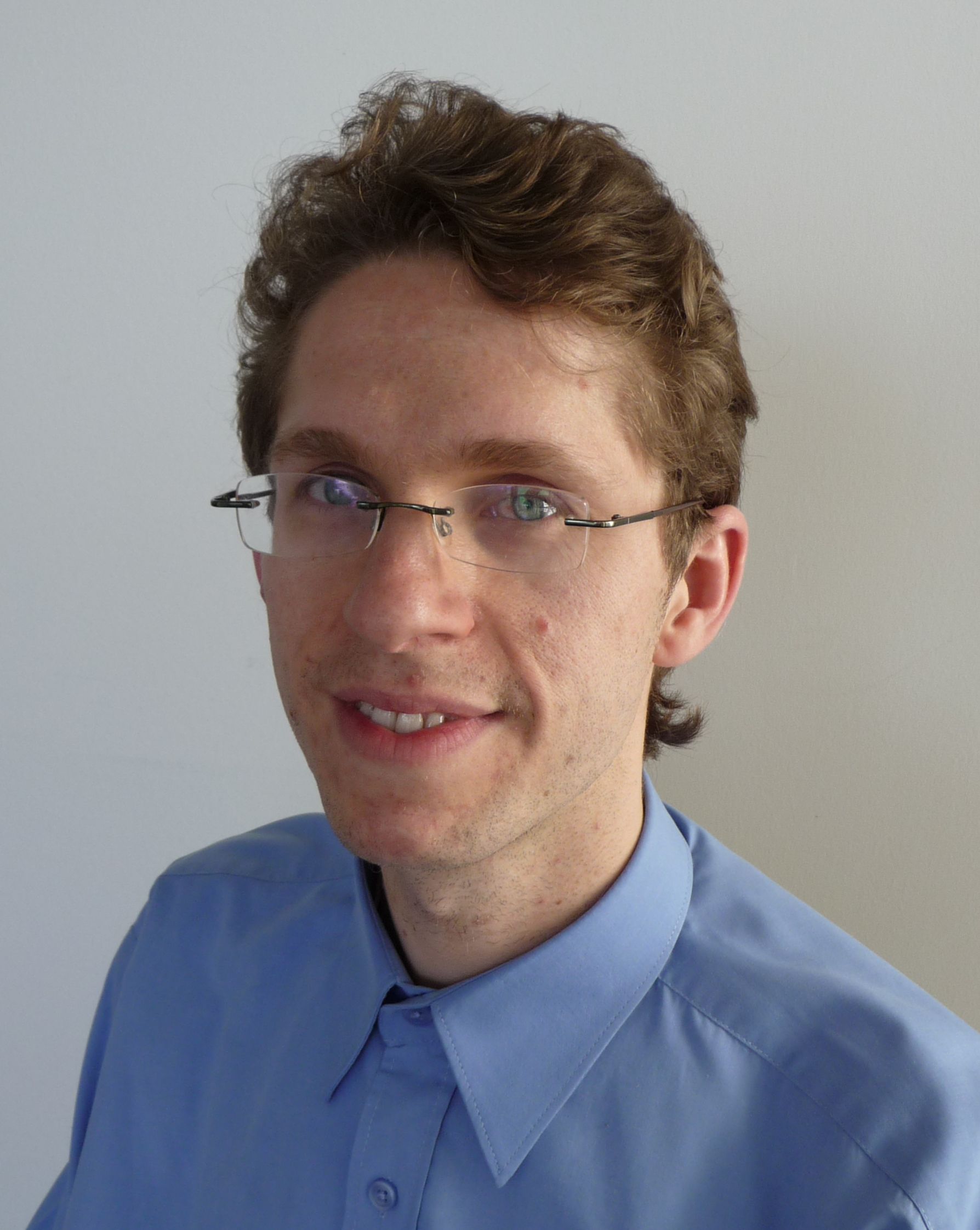}}
\textbf{David Schmitz} recieved the  Ph.D. in computer science from Ludwig Maximilians Universität (LMU) München in 2008. Since 2002 he is working at the Leibniz Supercomputing Center (LRZ) in Germany at the begining for the German Research Netwok and later for the European project Géant. He designed and developed the Customer Network Managemnt (CNM) tool. Curently he is working on the further development CNM tool and on the visualisation of the LHCOPN within the european Géant projekt.
\\
\\
\label{sec:WolfgangFritz}
\parpic[r]{\includegraphics[width=0.2\textwidth]{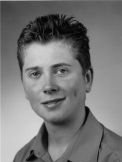}}
\textbf{Wolfgang Fritz} received diploma in Information Technology in 2009 from the department of Electrical Engineering at Technische Universität München. In late 2009, he joined the Leibniz Supercomputing Centre in Garching/Munich as a research assistant. His current research focusses on multi-domain circuit management and monitoring, especially within the European G\'eant project. Of special interest are not only technical, but also organizational approaches.
\\ 
\\ 

\label{sec:MarkYampolskiy}

\parpic[r]{\includegraphics[width=0.2\textwidth]{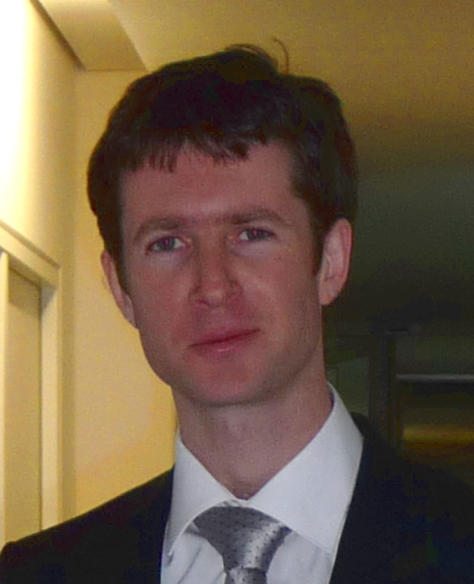}}
\textbf{Mark Yampolskiy} has a Ph.D. in computer science from Ludwig Maximilians Universität (LMU) München. Since 2006 he is working at the Leibniz Supercomputing Center (LRZ) in Germany for the European project Géant. He is also a member of the MNM team. His current research focuses on QoS assurance for point-to-point network connection and on the cryptographical aspects of IT security. 
\\
\\
\\

\label{sec:WolfgangHommel}
\parpic[r]{\includegraphics[width=0.2\textwidth]{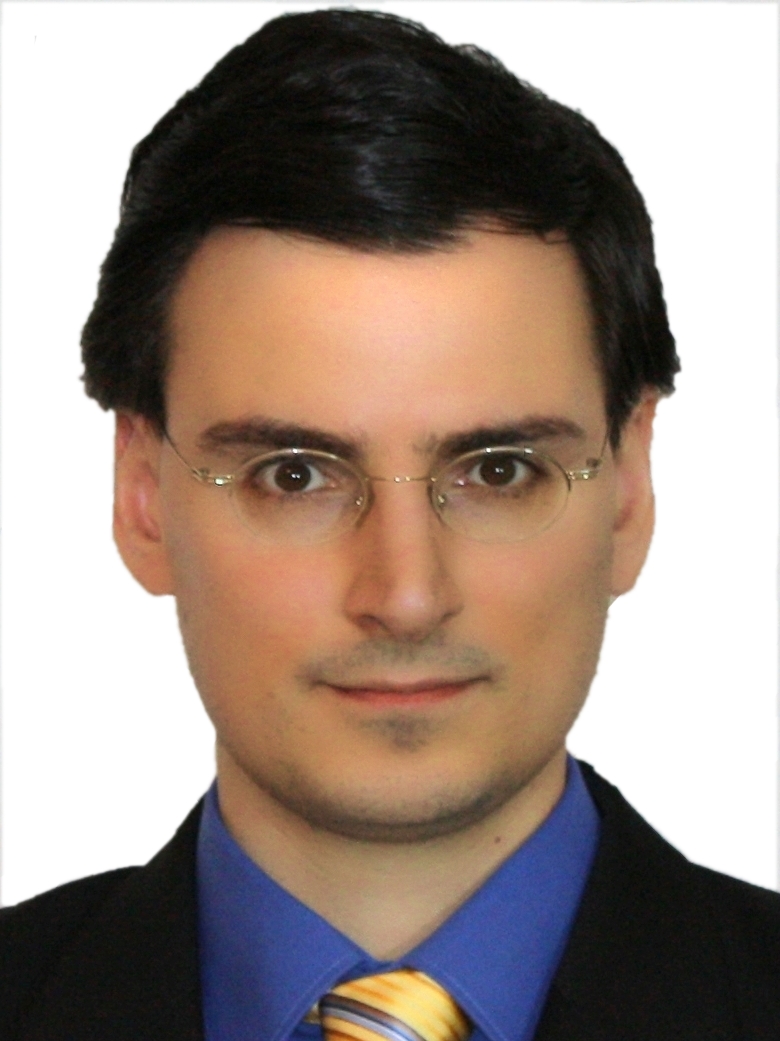}}

\textbf{Wolfgang Hommel} has a Ph.D. in computer science from Ludwig Maximilians Universität (LMU) München, and heads the network services planning group at the Leibniz Supercomputing Centre (LRZ) in Germany. His current research focuses on IT security and privacy management in large distributed systems, including identity federations and Grids. Emphasis is put on a holistic perspective, i.e. the problems and solutions are analyzed from the design phase through software engineering, deployment in heterogeneous infrastructures, and during the
operation and change phases according to IT service management process frameworks, such as ISO/IEC 20000-1. 

\end{document}